\documentclass[reprint,superscriptaddress,showpacs,amsmath,amssymb,aps,pra]{revtex4-2}
\usepackage{pifont}
\usepackage{times}
\usepackage{amssymb}
\usepackage{graphicx}
\usepackage{dcolumn}
\usepackage{dsfont}
\usepackage{bm}
\usepackage{subfigure}
\usepackage[ruled]{algorithm2e}
\usepackage{hyperref}
\usepackage{appendix}
\usepackage{lipsum}

\hypersetup{colorlinks=true, citecolor=blue, urlcolor=blue, linkcolor=blue}

\newcommand{\bra}[1]{\langle #1|}
\newcommand{\ket}[1]{|#1 \rangle}
\newcommand{\braket}[2]{ \langle #1| #2 \rangle}
\newcommand{\ketbra}[2]{|#1 \rangle\! \langle#2|}

\input amssym.def

\begin{document}
\title{Geometric-Like imaginarity: quantification and state conversion}
\author{Meng-Li Guo}
\affiliation{School of science, East China University of Technology, Nanchang 330006, China}
\author{Bo Li}
\email{libobeijing2008@163.com}
\affiliation{School of Computer and Computing Science, Hangzhou City University, Hangzhou 310015, China}
\author{Shao-Ming Fei}
\email{feishm@cnu.edu.cn}
\affiliation{School of Mathematical Sciences, Capital Normal University, Beijing 100048, China}
	
\begin{abstract}
From the perspective of resource-theoretic approach, this study explores the quantification of imaginary in quantum physics. We propose a well defined measure of imaginarity, the geometric-like measure of imaginarity. Compared with the usual geometric imaginarity measure, this geometric-like measure of imaginarity exhibits smaller decay difference under quantum noisy channels and higher stability. As applications, we show that both the optimal probability of state transformations from a pure state to an arbitrary mixed state via real operations, and the maximal probability of stochastic-approximate state transformations from a pure state to an arbitrary mixed state via real operations with a given fidelity $f$, are given by the geometric-like measure of imaginarity.
\end{abstract}
	
\maketitle
	
\section{Introduction}
A key feature of quantum mechanics is the need of imaginary numbers to accurately simulate the dynamics of physical systems. Although imaginary numbers have been long used to simplify the models of oscillatory motions and wave mechanics in classical physics, they seem to play a more deeper role in quantum physics as they are intrinsic to any orthodox formulation\cite{Wootters37,H38,Aleksandrova39}. For example, consider the polarization density matrix of a single photon in the $\{|H\rangle,\,|V\rangle\}$ basis, where $|H \rangle$ and $|V\rangle$ express the horizontal polarization and vertical polarization, respectively. The imaginary number in the density matrix causes the rotation of the electric field vector, that is, elliptical or circular polarization. According to the postulates of quantum mechanics, the state of a quantum system may be described by the wave function $\Psi(x)=|\Psi(x)| e^{-i\phi(x)}$, with the probability amplitude $|\Psi(x)|^2$ and the phase $\phi(x)$. It is natural to ask whether it is necessary to describe the fundamental properties and dynamics of a quantum system with the imaginary part $e^{-i\phi}$. That is to say, can quantum physics be formulated with only real numbers?
Recently, the answer to this question has been proven to be negative\cite{Wu1,Renou2,Chen3,Li4}.

Quantum resource theory provides a unified method in studying various quantum phenomena and their applications in quantum information protocols\cite{Chitambar51}. It provides an operational framework to quantify and manipulate the resources of quantum systems\cite{Chitambar51,Brandao1}. A resource theory consists two elements: the free states and the free operations. The free states are the quantum states with vanishing resources. And resource states can not be created from any free states through any free operations. A well-known example is the resource theory of quantum entanglement\cite{Horodecki1}. The consumption of entanglement leads to protocols such as superdense coding\cite{Bennett3} and quantum teleportation\cite{Bennett2}. Since then other quantum resource theories have been developed, such as resource theories of coherence\cite{Baumgratz4,Chitambar52,Marvian6,ChitambarHsieh7,Winter8,Napoli9}, athermality \cite{Brandao10,Horodecki11,Lostaglio12,Gour13,Narasimhachar14}, asymmetry\cite{Gour15,Gour16,Skotiniotis17,Marvian18}, knowledge\cite{Rio19}, magic\cite{Veitch20,Howard21,Ahmadi22}, superposition\cite{Theurer23}, steering\cite{Gallego33}, nonlocality\cite{Vicente34,LXB} and contextuality\cite{Amaral35}.
Recently, the theory of imaginative operational resources has been introduced\cite{Hickey1}. Then, Wu et al. further developed a more detailed theory of imaginarity resource, which was promoted by the latest development of entanglement and coherence theories\cite{Wu1,Wu17}. They studied the quantification of imaginarity based on two specific measures of imaginarity: geometric imaginarity and the robustness of imaginarity, and presented the operational explanations in state transformation problems. These works are not only of fundamental importance to the study of imaginarity resource theory, but also of operational significance.

A quantitative analysis of imaginarity contributes to the development of quantum technology such as quantum computers. In fact, different quantification of imaginarity can greatly enrich our understanding of imaginarity. Up to now only a few specific imaginarity measures have been introduced\cite{Gao,WuNew,Quantum}. In particular, the trace norm of imaginarity\cite{Hickey1,Wu1}, the fidelity of imaginarity\cite{Wu17,Kondra11}, the relative entropy of imaginarity\cite{Xue7}, the weight of imaginarity\cite{Xue7} and the imaginarity of bosonic Gaussian states\cite{Xu1} have been proposed and investigated.

In this paper, in the framework of resource theory we put forward a new imaginarity measure, the geometric-like imaginarity. By analyzing its basic properties and operational significance, and exploring its relationship with other imaginarity resource theories, we find that the geometric-like imaginarity measure has significant advantages compared with the geometric imaginarity measure. This study not only provides new theoretical support for the development of quantum technology, but also provides powerful tools for its practical application. The structure of the article is arranged as follows: the second part outlines the basic concepts of resource theory; the third section proposes the geometric-like imaginarity; the fourth part compares the decay value of geometric-like imaginarity and geometric imaginarity after quantum channel transmission; the fifth part applies the geometric-like imaginarity to the state conversion problem in the theory of virtual quantities and focuses on the optimal fidelity analytical expression of the conversion probability. Finally, the sixth part summarizes the main achievements and research significance.

\section{Preliminaries}
In the framework of quantum resource theory, a pivotal inquiry concerns the existence of a free operation, denoted as $\Phi_f$, that transforms a given quantum state $\rho$ into another state $\sigma$. Specifically, this transformation is expressed as $\sigma=\Phi_f [\rho]$\cite{Winter8}, indicating that $\rho$ possesses a greater resourcefulness compared to $\sigma$. For any resource measure $R$, the monotonicity principle holds,
\begin{eqnarray}
	R(\rho) \geq R(\sigma), \label{Monotonicity}
\end{eqnarray}
implying that $\rho$ contains at least as much resource content as $\sigma$.

In cases where a deterministic free operation cannot convert $\rho$ into $\sigma$ (e.g., when $R(\rho)<R(\sigma)$), probabilistic methods may still be applicable. This is feasible in resource theories that admit stochastic free operations, characterized by a set of free Kraus operators $\{K_j\}$  satisfying the condition $\sum_i K_j^\dagger K_j \leq \mathbb{I}$. Assume that any incomplete set of $\{K_j\}$ can be supplemented with additional operators $\{L_i\}$ to satisfy the completeness relation $\sum_i L_i^\dagger L_i + \sum_j K_j^\dagger K_j = \mathbb{I}$, the maximal conversion probability $P(\rho \rightarrow \sigma)$ is defined as
\begin{eqnarray}
	P\left(\rho\rightarrow \sigma\right)=\max_{\{K_j\}} \left\{\sum_j p_j: \sigma
	=\frac{\sum_j K_j \rho K_j^\dagger}{\sum_j p_j}\right\},\nonumber
\end{eqnarray}
where $p_j=tr[K_j\rho K_j^\dagger]$.
Notably, a deterministic operation exists when $P(\rho \rightarrow \sigma) = 1$. However, if both deterministic and stochastic conversions fail, leading to $P(\rho \rightarrow \sigma)=0$, approximate transformations may still be possible. The closeness of the achieved transformation to the desired one is quantified by the maximal transformation fidelity $F(\rho\rightarrow\sigma)$, defined as
\begin{eqnarray}
	F(\rho\rightarrow\sigma)=\max_{\Phi_f}\left\{F(\Phi_f[\rho],\sigma) \right\},\nonumber
\end{eqnarray}
where the maximum is taken over all permissible free operations $\Phi_f$ within the given resource theory.

Resource measures $R$ satisfy the monotonicity condition given by Eq.~(\ref{Monotonicity}). However, in stochastic theories, a stronger condition requires $R$ to be monotonic on average,
\begin{eqnarray}
	R(\rho) \geq \sum_j q_j R(\sigma_j),\nonumber 
\end{eqnarray}
where $\sigma_j = K_j \rho K_j^\dagger/q_j$ with free Kraus operators $K_j$, and $q_j$ is the corresponding probability, $q_j = tr[K_j \rho K_j^\dagger]$.
Resource quantifiers that satisfy this stronger condition are referred to ``strong resource monotones". Notably, if $R$ is also convex, strong monotonicity implies monotonicity.

Finally, for convex and strong monotonic resource quantifiers, a rigorous upper bound on the conversion probability can be formulated as
\begin{eqnarray}
	P(\rho \rightarrow \sigma) \leq \min\left\{ \frac{R(\rho)}{R(\sigma)}, 1 \right\},\label{conversion}
\end{eqnarray}
where the upper bound is given by the ratio of the resource quantities and the probability does not exceed unity.

Let $\{|j\rangle \}^{d-1} _ {j = 0} $ be a fixed base in $d$-dimensional Hilbert space $\mathcal{H}$. We denote $\mathcal{D}(\mathcal{H})$ the set of density operators acting on $\mathcal{H}$. In the context of the resource theory of imaginarity, the free states are the real ones defined by
\begin{eqnarray}
	\mathcal{F}=\{\rho \in \mathcal{D}(\mathcal{H}): \langle i|\rho|j \rangle \in \mathbb{R}, i,j=0,1,...,d-1\},\nonumber
\end{eqnarray}
where $\mathbb{R}$ represents the set of real numbers. Real operations are characterized by real Kraus operators. Specifically, an operation $\Phi$ characterized by Kraus operators $\{K_l\}$ is deemed real if $\langle i|K_l|j \rangle \in \mathbb{R}$ for any $l$ and $i,j=0,1,...,d-1$. A real-valued functional $M$ of quantum states is called an imaginarity measure if it satisfies the following conditions (M1) to (M4)\cite{Wu1,Wu17,Hickey1},

(M1) $Nonnegativity:$ $M(\rho )\geq 0$ and $M(\rho )=0$ if and only if $\rho$ is a real state.

(M2) $Monotonicity:$ $M[\Phi(\rho)]\leq M(\rho)$ if $\Phi$ is a real operation.

(M3) $Strong~imaginarity~monotonicity:$ $\sum_j p_j M(\rho_j)\leq M(\rho)$, where $p_j=\mathrm{Tr}K_j\rho K^{\dagger}_j$, $\rho_j= K_j\rho K^{\dagger}_j/p_j$ and $K_j$ are real Kraus operators.

(M4) $Convexity:$ $M(\sum_{j}p_{j}\rho _{j})\leq \sum_{j}p_{j}M(\rho _{j})$
for any probability distribution $\{p_{j}\}$ and states $\{\rho_{j}\}.$

\section{Geometric-Like imaginarity}\label{sec:3}
For a pure state $\ket{\psi}$, the geometric imaginarity is defined as\cite{Wu17}
\begin{eqnarray}
	\mathcal{M}_g (\ket{\psi})=1 - \max_{\ket{\phi} \in \mathcal{F}} |\braket{\phi}{\psi}|^2,\nonumber
\end{eqnarray}
where the maximization goes over all real pure states.
The geometric imaginarity $\mathcal{M}_g(\rho)$ of a mixed state $\rho$ is given by
\begin{eqnarray}
	\mathcal{M}_g(\rho)=\min \sum_j p_j\mathcal{M}_g (\ket{\psi_j}),\nonumber
\end{eqnarray}
where the minimum is taken over all ensembles $\{p_j,\ket{\psi_j}\}$ that satisfy $\rho = \sum_j p_j \ket{\psi_j}\!\bra{\psi_j}$. This definition of geometric imaginarity is equivalent to the following distance-based measure \cite{Streltsov34},
\begin{eqnarray}
	\mathcal{M}_g(\rho)=1-\max_{\sigma\in \mathcal{F}} F(\rho,\sigma)= \frac{1 - \sqrt{F(\rho,\rho^T)}}{2},\nonumber
\end{eqnarray}
where $\sqrt{F(\rho,\sigma)} =\mathrm{Tr}\sqrt{\sqrt{\rho} \sigma \sqrt{\rho}}$ represents the root fidelity.

Specifically, for any pure state $\ket{\psi}$, there exists a real orthogonal matrix $O$ such that
\begin{eqnarray}
	O\ket{\psi} = \sqrt{\frac{1+|\braket{\psi^*}{\psi}|}{2}}\ket{0} + i \sqrt{\frac{1-|\braket{\psi^*}{\psi}|}{2}}\ket{1}. \label{OOO}
\end{eqnarray}
Hence, for any pure state $\ket{\psi}$, any imaginarity measure $\mathcal{M}$ can be written as
\begin{eqnarray}
	\mathcal{M}(\ket{\psi}) = f(|\braket{\psi^*}{\psi}|)\label{PureMeasures}
\end{eqnarray}
for some suitable function $f$. For instance, the geometric imaginarity can be expressed as \cite{Wu17},
\begin{eqnarray}
	\mathcal{M}_g (\ket{\psi})=\frac{1-|\braket{\psi^*}{\psi}|}{2}\label{gmtu}.
\end{eqnarray}

In this study, we introduce a new imaginarity measure named geometric-like imaginarity.

{\bf Definition 1}.\label{lgpure}
The geometric-like imaginarity of any pure state $\ket{\psi}$ is defined by
\begin{eqnarray}
	\mathcal{M}_{gl}(\ket{\psi})=1-\sqrt{\frac{1+|\braket{\psi^*}{\psi}|}{2}}.\label{IM2}
\end{eqnarray}
The geometric-like imaginarity of a mixed state $\rho$ is defined via the convex-roof extension,
\begin{eqnarray}\label{glmixed}
	\mathcal{M}_{gl}(\rho)=\min_{\{p_j,\ket{\psi_j}\}}\sum_{j} p_j  \mathcal{M}_{gl}(\ket{\psi_j}),
\end{eqnarray}
where the minimum is taken over all the pure state decompositions of $\rho=\sum_j p_j \ketbra{\psi_j}{\psi_j}$ with $\sum_j p_j=1$ and $p_i\geq 0$.

Given that $\mathcal{M}_g$ is a convex function, we can deduce that $\mathcal{M}_{gl}$ is also convex,
\begin{eqnarray}\label{GLvex}
	\mathcal{M}_{gl}\Big(\sum_j p_j \rho_j\Big)\leq\sum_j p_j M_{gl}(\rho_j).
\end{eqnarray}
We have the following conclusion.

{\bf Theorem 1}.
$\mathcal{M}_{gl}(\rho)$ is a well-defined measure of imaginarity.

{\sf Proof.}
It is straightforward to observe that $\mathcal{M}_{gl}(\rho)=0$ if and only if $\rho$ is an real state. Thus the condition (M1) is fulfilled. Based on (\ref{GLvex}), we get that $\mathcal{M}_{gl}(\rho)$ also satisfies (M4).

To prove (M3), we first consider pure states. According to Eq. (\ref{OOO}), any dimensional pure states can be transformed into single-qubit states under a real orthogonal transformation. Since the imaginarity remains unchanged under the real orthogonal transformations,
let us evaluate $\mathcal{M}_{gl}$ for any state of the form $\ket{u} = x_0\ket{0} + ix_1\ket{1}$ with $x_0^2 + x_1^2 = 1$ and $ x_0^2\geq x_1^2\geq 0$. For any real state $\ket{v} =y_0\ket{0}+y_1\ket{1}$ with $y_0^2 + y_1^2 = 1$, we have
\begin{eqnarray}
	|\braket{v}{u}| =|x_0y_0 + ix_1y_1| \leq \sqrt{x_0^2}=|x_0|,\nonumber
\end{eqnarray}
where the inequality is due to $\sum_j y_j^2 = 1$ and $x_0 \geq x_1$. Since $|\braket{0}{u}|=|x_0|$ \cite{Wu17}, we conclude that
\begin{eqnarray}
	|x_0|=\max_{\ket{v} \in \mathcal{F}} |\braket{v}{u}|.\nonumber
\end{eqnarray}
Consequently,
\begin{eqnarray}
	\mathcal{M}_{gl}(\ket{\psi})&=&1-\sqrt{\frac{1+|\braket{\psi^*}{\psi}|}{2}}\nonumber\\
	&=&1-\max_{\ket{\phi}\in \mathcal{F}}|\braket{\phi}{\psi}|.\nonumber
\end{eqnarray}
Without loss of generality, we consider states $\ket{\theta} = \cos\theta \ket{0} + i \sin\theta \ket{1}$ with $0\leq \theta \leq \frac{\pi}{4}$, for which the geometric-like imaginarity is $\mathcal{M}_{gl}(\ket{\theta})=1-\cos\theta$. As all post-measurement states are remain pure for pure initial states, proving (M3) for pure states reduces to prove the following inequality,
\begin{eqnarray}\label{cos1}
	\sum_j\max_{\ket{\nu_j}\in\mathcal{F}} |\braket{\nu_j}{K_j|\theta}| \geq \cos\theta,
\end{eqnarray}
where $\{K_j\}$ represents a set of real Kraus operators. To prove (\ref{cos1}), let $\ket{\nu_j}=\frac{K_j\ket{0}}{\sqrt{m_j}}$ with $m_j=\braket{0}{K_j^T K_j|0}$. We have
\begin{eqnarray}
	\sum_j\max_{\ket{\nu_j}\in\mathcal{F}} |\braket{\nu_j}{K_j|\theta}| \geq \sum_j \frac{|\braket{0}{K_j^T K_j|\theta}|}{s_j}.\nonumber
\end{eqnarray}
Since all Kraus operators $K_j$ are real, we obtain
\begin{align}
	|\braket{0}{K_j^T K_j|\theta}|=&| \cos \theta \braket{0}{K_j^T K_j|0}+i\sin \theta \braket{0}{K_j^T K_j|1}|  \nonumber \\
	\geq&\cos \theta |\braket{0}{K_j^T K_j|0}|. \nonumber
\end{align}
This implies that
\begin{eqnarray}
	\sum_j\max_{\ket{\nu_j}\in\mathcal{F}} |\braket{\nu_j}{K_j|\theta}| \geq \sum_j \frac{|\braket{0}{K_j^T K_j|0}|}{m_j}\cos \theta.\nonumber
\end{eqnarray}
Taking into account the identity $\sum_j K_j^T K_j =\mathcal{I}$, we prove the inequality (\ref{cos1}).

Therefore $\mathcal{M}_{gl}$ satisfies (M3) for pure states. To extend the conclusion to mixed states, we consider an optimal decomposition of a mixed state $\rho=\sum_j p_j \ket{\psi_j}\!\bra{\psi_j}$ such that
\begin{eqnarray}
	\mathcal{M}_{gl}(\rho)=\sum_j p_j \mathcal{M}_{gl}(\ket{\psi_j}).\nonumber
\end{eqnarray}
Set $m_{jk}=\braket{\psi_k}{K_j^T K_j|\psi_k}$. Using the convexity of $\mathcal{M}_{gl}$, we obtain
\begin{align}
	\sum_j q_j \mathcal{M}_{gl}\left(\frac{K_j\rho K_j^T}{q_j}\right) 
	&=\sum_j q_j \mathcal{M}_{gl}\left(\sum_k p_k \frac{K_j \ket{\psi_k}\!\bra{\psi_k} K_j^T}{q_j}\right) \nonumber \\
	&=\sum_j q_j \mathcal{M}_{gl}\left(\sum_k \frac{p_k m_{jk}}{q_j} \times \frac{K_j \ket{\psi_k}\!\bra{\psi_k} K_j^T}{m_{jk}}\right) \nonumber \\
	&\leq \sum_{j,k} p_k m_{jk} \mathcal{M}_{gl}\left( \frac{K_j \ket{\psi_k}\!\bra{\psi_k} K_j^T}{m_{jk}}\right) \nonumber \\
	&\leq \sum_j p_j \mathcal{M}_{gl}(\ket{\psi_j}) \nonumber \\
	&= \mathcal{M}_{gl}(\rho).\nonumber
\end{align}
This completes the proof that (M3) holds for all mixed states. 
Monotonicity follows directly from the convexity of $\mathcal{M}_{gl}(\rho)$, $\mathcal{M}_{gl}(\rho)\geq \mathcal{M}_{gl}(\sum_{j}p_j\rho_j)=\mathcal{M}_{gl}(\Phi(\rho))$, thus $\mathcal{M}_{gl}(\rho)$ also satisfies (M2).
$\hfill\blacksquare$

The geometric-like measure of imaginarity has the following analytical expression.

{\bf Theorem 2}.\label{Thm:geom_meas}
The geometric-like imaginarity of an arbitrary quantum state $\rho$ is given by
\begin{eqnarray}
	\mathcal{M}_{gl}(\rho) =1-\sqrt{ \frac{1+\sqrt{F(\rho,\rho^T)}}{2}}, \label{geoimexp}
\end{eqnarray}
where $F(\rho,\rho^T)$ is the fidelity between $\rho$ and its transpose $\rho^T$.

{\sf Proof.}
For any mixed state $\rho$ with pure state decomposition $\rho=\sum_j p_j \ket{\psi_j}\!\bra{\psi_j}$, it has been shown in Ref.~\cite{Kondra11} that
\begin{eqnarray}\label{opt_mod}
	\max_{\{p_j,\ket{\psi_j}\}}\sum_{j} p_j|\braket{\psi^*_j}{\psi_j}|\leq \sqrt{F(\rho,\rho^T)}.
\end{eqnarray}
Therefore, for any pure state decomposition of $\rho$ we have
\begin{eqnarray}
	\mathcal{M}_{gl}(\rho)&=&\min_{\{p_j,\ket{\psi_j}\}}\sum_{j} p_j\Bigg(1-\sqrt{\frac{1+|\braket{\psi_j^*}{\psi_j}|}{2}}\Bigg)\nonumber\\
	&\geq&\min_{\{p_j,\ket{\psi_j}\}}\Bigg(1-\sqrt{\frac{1+\sum_{j} p_j|\braket{\psi_j^*}{\psi_j}|}{2}}\Bigg)\nonumber\\
	&\geq& 1-\sqrt{\frac{1+\sqrt{F(\rho,\rho^T)}}{2}}.
\end{eqnarray}
To demonstrate that the equality in the above inequality holds, we construct a pure state decomposition to attain the lower bound. As outlined in Ref.~\cite{Kondra11}, such a decomposition can be formulated via singular value decomposition. Let $\rho=\sum_k \lambda_k \ket{\mu_k}\!\bra{\mu_k}$ be the spectral decomposition of $\rho$, where $\ket{\mu_k}$ are the eigenstates of $\rho$ and $\lambda_k$ are the corresponding eigenvalues. Given that $\sqrt{F(\rho,\rho^T)}$ relates to the eigenvalues of $\sqrt{\sqrt{\rho}  \rho^{T}\sqrt{\rho}}$, we employ a methodology akin to Ref.~\cite{Kondra11} to select $\ket{\mu_k}$ such that $|\braket{\mu_i}{\mu^{*}_j}|$ equals the square root of the corresponding eigenvalue of $\sqrt{\sqrt{\rho}  \rho^{T}\sqrt{\rho}}$. Consequently, we construct a pure state decomposition such that
\begin{eqnarray}
	\mathcal{M}_{gl}(\rho)=1-\sqrt{\frac{1+\sqrt{F(\rho,\rho^T)}}{2}}.\nonumber
\end{eqnarray}
$\hfill\blacksquare$

\section{Decay of imaginarity of pure states under quantum channels}\label{sec:4}
Quantum systems are inherently influenced by their environments. Similar to quantum entanglement and coherence, the imaginarity of a quantum state will also be affected by the environments. The impact of noisy environments on the imaginarity of quantum states needs to be analyzed. This section aims to analyze the geometric imaginarity and the geometric-like imaginarity of pure states under some quantum channels. Specifically, we delve into how quantum noise affects these imaginarities, analogous to the studies on coherence decay \cite{Xi56}.

For a general quantum channel $\varepsilon$, the decay of the imaginarity can be defined as the difference between the initial state's imaginarity and that after the application of the noisy channel, i.e., $\Delta\mathcal{M}=\mathcal{M}(\rho)-\mathcal{M}(\varepsilon(\rho)).$
According to (\ref{OOO}) and fact that the imaginarity of any pure state $\rho$ remains unchanged under the real orthogonal transformation $O$, $\mathcal{M}(O\rho O^T)=\mathcal{M}(\rho)$, we consider the following general form,
\begin{eqnarray}
	\ket{\eta}&=&\sqrt{\frac{1+|\braket{\psi^*}{\psi}|}{2}}\ket{0} + i \sqrt{\frac{1-|\braket{\psi^*}{\psi}|}{2}}\ket{1}\nonumber\\
	&=&\sqrt{\frac{1+A}{2}}\ket{0} + i \sqrt{\frac{1-A}{2}}\ket{1},\nonumber
\end{eqnarray}
where $A=|\braket{\psi^*}{\psi}|$.

Consider the bit flip (BF), phase damping (PD) and amplitude damping (AD) channels given by the following Kraus operators, respectively,
\begin{align}
	E^{BF}_0&=\left(
	\begin{array}{cc}
		\sqrt{m} & 0 \\
		0 & \sqrt{m} \\
	\end{array}
	\right),~~E^{BF}_1= \left(
	\begin{array}{cc}
		0 & \sqrt{1-m} \\
		\sqrt{1-m} & 0 \\
	\end{array}
	\right);\nonumber\\
	E^{PD}_0&=\left(
	\begin{array}{cc}
		1 & 0 \\
		0 & \sqrt{1-n} \\
	\end{array}
	\right),~~E^{PD}_1=\left(
	\begin{array}{cc}
		0 & 0 \\
		0 & \sqrt{n} \\
	\end{array}
	\right);\nonumber\\
	E^{AD}_0&=\left(
	\begin{array}{cc}
		1 & 0 \\
		0 & \sqrt{1-p} \\
	\end{array}
	\right),~~E^{AD}_1=\left(
	\begin{array}{cc}
		0 & \sqrt{p} \\
		0 & 0 \\
	\end{array}
	\right).\nonumber
\end{align}
Under the BF, PD and AD channels, a state $\rho$ is transformed into, respectively,
\begin{eqnarray}
	\rho_{BF}=\left(
	\begin{array}{cc}
		A \left(m-\frac{1}{2}\right)+\frac{1}{2} & \frac{1}{2} i \sqrt{1-A^2} (1-2 m) \\
		\frac{1}{2} i \sqrt{1-A^2} (2 m-1) & \frac{1}{2} (-2 A m+A+1) \\
	\end{array}
	\right),\nonumber
\end{eqnarray}
\begin{eqnarray}
	\rho_{PD}=\left(
	\begin{array}{cc}
		\frac{A+1}{2} & -\frac{1}{2} i \sqrt{1-A^2} \sqrt{1-n} \\
		\frac{1}{2} i \sqrt{1-A^2} \sqrt{1-n} & \frac{1-A}{2} \\
	\end{array}
	\right),\nonumber
\end{eqnarray}
\begin{eqnarray}
	\rho_{AD}=\left(
	\begin{array}{cc}
		\frac{1}{2} (A (-p)+A+p+1) & -\frac{1}{2} i \sqrt{1-A^2} \sqrt{1-p} \\
		\frac{1}{2} i \sqrt{1-A^2} \sqrt{1-p} & \frac{1}{2} (A-1) (p-1) \\
	\end{array}
	\right).\nonumber
\end{eqnarray}
Through meticulous algebraic manipulations, we obtain the following expressions for the decay of geometric-like and geometric imaginarity under these channels,
\begin{eqnarray}
	\Delta\mathcal{M}^{BF}_{gl}&=&\frac{2 \sqrt{\frac{\sqrt{s_1-t_1}+\sqrt{s_1+t_1}}{\sqrt{2}}+1}-2 \sqrt{A+1}}{2 \sqrt{2}},\nonumber\\
	\Delta\mathcal{M}^{PD}_{gl}&=&\frac{1}{2} \Big[\sqrt{\sqrt{s_2-t_2}+\sqrt{s_2+t_2}+2}-\sqrt{2} \sqrt{A+1}\Big],\nonumber\\
	\Delta\mathcal{M}^{AD}_{gl}&=&\frac{1}{2} \Big[\sqrt{\sqrt{s_3-t_3}+\sqrt{s_3+t_3}+2}-\sqrt{2} \sqrt{A+1}\Big],\nonumber
\end{eqnarray}

\begin{widetext}
	\begin{figure*}[tb]
	\includegraphics[width=18cm]{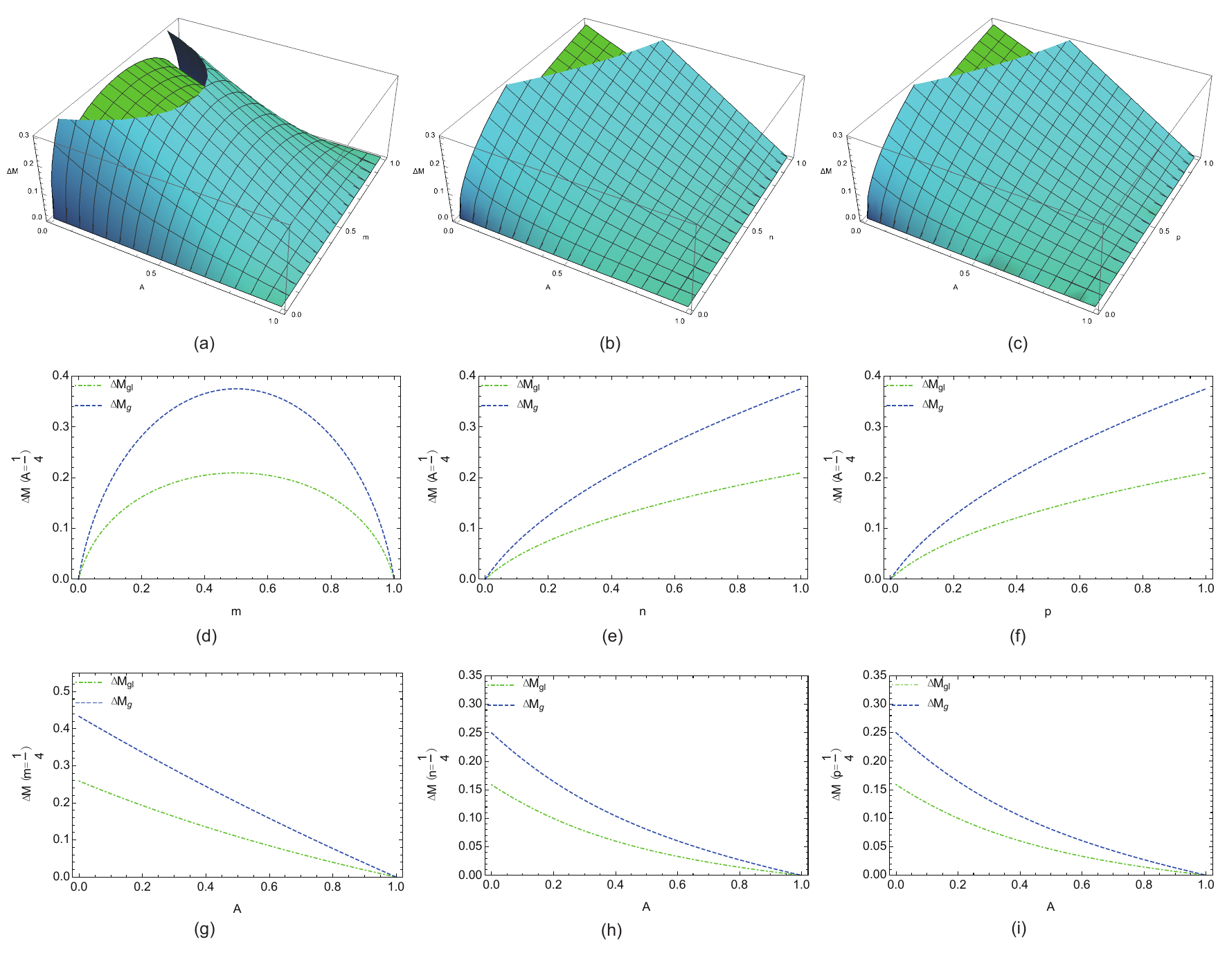}
	\caption{The decay of $\Delta\mathcal{M}_{gl}$ and $\Delta\mathcal{M}_{g}$ with respect to different channels: (a), (d) and (g) represent bit flip channel attenuation for $\Delta\mathcal{M}_{gl}$ and $\Delta\mathcal{M}_{g}$, while (b), (e) and (h) correspond to phase damping channel attenuation for $\Delta\mathcal{M}_{gl}$ and $\Delta\mathcal{M}_{g}$, and (c), (f) and (i) are for the amplitude damping channel attenuation for $\Delta\mathcal{M}_{gl}$ and $\Delta\mathcal{M}_{g}$. The blue surface in (a), (b) and (c) denotes $\Delta\mathcal{M}_{g}$, while the green surface represents $\Delta\mathcal{M}_{gl}$.}
	\label{Fig:1}
	\end{figure*}
\end{widetext}

and
\begin{eqnarray}
	\Delta\mathcal{M}^{BF}_{g}&=&\frac{1}{4} \Big[\sqrt{2} \left(\sqrt{s_1-t_1}+\sqrt{s_1+t_1}\right)-2 A\Big],\nonumber\\
	\Delta\mathcal{M}^{PD}_{g}&=&-\frac{A}{2}+\frac{\sqrt{s_2-t_2}}{4}+\frac{\sqrt{s_2+t_2}}{4},\nonumber\\
	\Delta\mathcal{M}^{AD}_{g}&=&\frac{1}{4}\Big(-2 A+\sqrt{s_3-t_3}+\sqrt{s_3+t_3}\Big),\nonumber
\end{eqnarray}
where $s_1=A^2 (1-2 m)^2-2 (m-1) m$, $t_1=A (1-2 m) \sqrt{A^2 (1-2 m)^2-4 (m-1) m}$, $s_2=n-A^2 (n-2)$, $t_2=2 \sqrt{A^4 (-n)+A^2 n+A^4}$, $s_3=A^2 (p-2) (p-1)-2 A (p-1) p+p^2+p$ and $t_3=2 \sqrt{\left(p-A^2 (p-1)\right) (A (-p)+A+p)^2}$.

The decay functions $\Delta\mathcal{M}_{gl}$ and $\Delta\mathcal{M}_{g}$ are shown in Figure \ref{Fig:1} for various scenarios. Our analysis reveals that both $\Delta\mathcal{M}_{gl}$ and $\Delta\mathcal{M}_{g}$ achieve their maximum values for maximally imaginary states ($A=0$) and vanish for real states ($A=1$). For a fixed quantum state (fixed $A$), the decay functions exhibit concavity with respect to the channel parameters $m,n,p$. Conversely, for a fixed channel (fixed $m,n,p$), the decay functions are convex and decreasing with respect to $A$.

It can be observed from the Figure \ref{Fig:1} that $\Delta\mathcal{M}_{g}>\Delta\mathcal{M}_{gl}$. Furthermore, the graph indicates that geometric imaginarity decreases rapidly (blue graphs), while geometric-like imaginarity (green graphs) demonstrates greater stability under the quantum channel. Therefore, under the quantum channel $\epsilon$, the geometric-like imaginarity exhibits a smaller attenuation difference, which means that it loses less information during transmission and has higher stability.

\section{State transformations via real operation} \label{sec:5}
In the following, we demonstrate the pivotal role played by the geometric-like measure of imaginarity in probabilistic and stochastic-approximate state transformations.

\subsection{Probabilistic transformations}
We now present the maximum probability of transforming a pure state $\ket{\psi}$ into another pure state $\ket{\phi}$ through real operations.

{\bf Proposition 1}. \label{PureConversion}
The maximum probability $P(\,\ket{\psi} \rightarrow \ket{\phi}\,)$ for a pure state transformation $\ket{\psi} \rightarrow \ket{\phi}$ via real operations is given by
\begin{eqnarray}
	P(\,\ket{\psi} \rightarrow \ket{\phi}\,) = \min \left\{~\frac{\sqrt{2}-\sqrt{1 +|\braket{\psi^*}{\psi}|}}{\sqrt{2}-\sqrt{1 +|\braket{\phi^*}{\phi}|}},~1~\right\}.\label{probpure}
\end{eqnarray}

{\sf Proof.}
From previous results \cite{Wu17}, we know that for any states $\rho$ and $\sigma$,  
\begin{eqnarray}  
	P(\rho \rightarrow \sigma) \leq \frac{\mathcal{M}_{gl}(\rho)}{\mathcal{M}_{gl}(\sigma)}.\nonumber
\end{eqnarray}
For pure states $\ket{\psi}$ and $\ket{\phi}$, this implies  
\begin{eqnarray}  
	P(\ket{\psi} \rightarrow \ket{\phi}) \leq \frac{\sqrt{2} - \sqrt{1 + |\braket{\psi}{\psi^\ast}|}}{\sqrt{2} - \sqrt{1 + |\braket{\phi}{\phi^\ast}|}}.\nonumber
\end{eqnarray}  

Next, consider the scenario where $|\braket{\psi}{\psi^\ast}| \geq |\braket{\phi}{\phi^\ast}|$. Define the state  
\begin{eqnarray}  
	\ket{\psi'} = \sqrt{\frac{1 + |\braket{\psi}{\psi^\ast}|}{2}}\ket{0} + i \sqrt{\frac{1 - |\braket{\psi}{\psi^\ast}|}{2}}\ket{1}.\nonumber
\end{eqnarray}  
Using real operations with Kraus operators \cite{Wu17}  
\begin{eqnarray}  
	K_0 = \begin{pmatrix}  
		a & 0 \\  
		0 & 1  
	\end{pmatrix}, \quad K_1 = \sqrt{\mathbb{I} - K_0^2},  \nonumber
\end{eqnarray}  
where $a \leq 1$,  we transform $\ket{\psi'}$ into  
\begin{eqnarray}  
	\ket{\phi'} = \sqrt{\frac{1 + |\braket{\phi}{\phi^\ast}|}{2}}\ket{0} + i \sqrt{\frac{1 - |\braket{\phi}{\phi^\ast}|}{2}}\ket{1},  \nonumber
\end{eqnarray}  
with probability  
\begin{eqnarray}  
	p = \frac{\sqrt{2} - \sqrt{1 + |\braket{\psi}{\psi^\ast}|}}{\sqrt{2} - \sqrt{1 + |\braket{\phi}{\phi^\ast}|}}.  \nonumber
\end{eqnarray}  
Since $\ket{\phi'}$ is equivalent to $\ket{\phi}$ up to a real orthogonal transformation, this probability is achievable.  

For $|\braket{\psi}{\psi^\ast}| < |\braket{\phi}{\phi^\ast}|$, the transformation $\ket{\psi} \rightarrow \ket{\phi}$ can be achieved with unit probability \cite{Hickey1}. Therefore, the maximum probability is as stated.  
$\hfill\blacksquare$

As exact deterministic state transformations are not always feasible, we consider the probabilistic scenario. The probabilistic transformations between pure states have been investigated in \cite{Wu1,Wu17}. In the theorem below, we extend this results to the case that the target state is mixed.

{\bf Theorem 3}.\label{333}
The optimal probability $P(\ket{\psi} \rightarrow \rho )$ of transforming a pure state $\ket{\psi}$ into a mixed state $\rho$ via real operations is given by
\begin{eqnarray} \label{optprob}
	P(\ket{\psi} \rightarrow \rho ) = \min\Biggl\{~\frac{\mathcal{M}_{gl}(\ket{\psi})}{\mathcal{M}_{gl} (\rho)},~1~\Biggr\}.
\end{eqnarray}

{\sf Proof.}
From (\ref{conversion}) we see that the ratio of geometric-like measures provides an upper bound on the optimal achievable probability $P(\sigma \rightarrow \rho )$ of transforming a state $\sigma$ to $\rho$,
\begin{eqnarray}
	P(\sigma \rightarrow \rho ) \leq \min\Biggl\{~\frac{\mathcal{M}_{gl}(\sigma)}{\mathcal{M}_{gl} (\rho)},~1~\Biggl\}.\label{probstrat}
\end{eqnarray}
Proposition 1 demonstrates that this inequality is saturated for probabilistic pure-to-pure state transformations, where the optimal probability is given by the ratio of geometric-like measures as shown in (\ref{probpure}).

We now prove that (\ref{probstrat}) is saturated when $\sigma=\ket{\psi}\bra{\psi}$ is a pure state. Let $\{p'_j,\ket{\psi'_j}\}$ be the optimal ensemble such that 
$\mathcal{M}_{gl}(\rho)=\sum_{j} p'_j \mathcal{M}_{gl}(\ket{\psi'_j})$.
We adopt a specific purification of $\rho$ \cite{Kondra11},
\begin{eqnarray}
	\ket{\rho} = \sum_j \sqrt{p_j'}\ket{\psi_j'} \otimes \ket{j}^{A},\nonumber
\end{eqnarray}
where $A$ signifies an ancillary system for purification. It can be easily verified that
$\mathcal{M}_{gl}(\ket{\rho}) =\mathcal{M}_{gl}(\rho)$. From (\ref{probpure}) we have
\begin{eqnarray*}
	P(\ket{\psi} \rightarrow \ket{\rho} )
	&&=\min\Biggl\{~\frac{\mathcal{M}_{gl}(\ket{\psi})}{\mathcal{M}_{gl} (\ket{\rho})},~1~\Biggl\}\nonumber\\
	&&=\min\Biggl\{\frac{\mathcal{M}_{gl}(\ket{\psi})}{\mathcal{M}_{gl} (\rho)},1\Biggl\}.
\end{eqnarray*}
By discarding the ancilla we complete the proof.
$\hfill\blacksquare$

Theorem 3 shows that the optimal probability of converting a pure state into a mixed state under real operations is given by the readily computable geometric-like measure of imaginarity.

\subsection{Stochastic-approximate transformations}
Next we extend our analysis to the framework that encompasses both stochastic and approximate scenarios. Stochastic-approximate state transformations have attracted considerable interest within the realm of general resource theories \cite{Regula25,Kondra29,Regula30,Fang35}. Specifically, the probability of stochastic-approximate conversion is defined as the maximum likelihood of transforming a state $\rho$ into $\sigma$ with fidelity at least $f$ \cite{Kondra29},
$$
P_{f}(\rho\rightarrow\sigma)=\max_{\Phi}\left\{\mathrm{tr}\Phi(\rho) : F\left(\frac{\Phi(\rho)}{\mathrm{tr}[ \Phi(\rho)]},\sigma\right)\geq f\right\},
$$
where $\Phi$ represents the set of all real operations. Similarly, the fidelity for stochastic-approximate conversion quantifies the maximal fidelity achievable in transforming a state $\rho$ into $\sigma$ with a success probability at least $p$ \cite{Kondra29},
$$
F_{p}(\rho\rightarrow\sigma)=\max_{\Phi}\left\{F\left(\frac{\Phi(\rho)}{\mathrm{tr}[ \Phi(\rho)]},\sigma\right):\mathrm{tr}\Phi(\rho)\geq p \right\}.
$$
Before delving into our main results, we provide the following lemmas.

{\bf Lemma 1}. \label{equal}
For any state $\rho$, there exists a pure state decomposition $\{p_i,\ket{\psi_i}\}$ such that $\mathcal{M}_{gl}(\ket{\psi_i})=\mathcal{M}_{gl}(\rho)$ for all $i$.

{\sf Proof.}
Given a pure state decomposition $\{\lambda_j,\ket{\mu_j}\}$ satisfying $\sqrt{\lambda_i \lambda_j}\braket{\mu_i}{\mu^{*}_j} = \delta_{ij}D_{j}$, where $D_j$ is the $j$th eigenvalue of $\sqrt{\sqrt{\rho}  \rho^{T}\sqrt{\rho}}$ \cite{Kondra11}, we have $\mathcal{M}_{gl}(\rho)=\sum_j\lambda_j\mathcal{M}_{gl}(\ket{\mu_j})$. We aim to achieve a decomposition $\{p_i, \ket{\psi_i}\}$ such that
$\braket{\psi_i}{\psi^{*}_i}= \braket{\psi_j}{\psi^{*}_j}$ $\forall\, i,j$. 
Note that $\braket{\mu_i}{\mu^{*}_i} \geq 0$ $\forall\, i$. 
We have $\sum_i \lambda_i \braket{\mu_i}{\mu^{*}_i}=1-\mathcal{M}_{gl}(\rho)$.
If $\mathcal{M}_{gl}(\ket{\mu_i})=\mathcal{M}_{gl}(\rho)$ for all $i$, we have the desired decomposition. We seek new pure state decompositions $\{\lambda_j',\ket{\mu_j'}\}$ such that $\sum_i \lambda_i' \braket{\mu_i'}{\mu'^{*}_i}=1-\mathcal{M}_{gl}(\rho)$ is kept.

Assume that two pure states $\ket{\mu_1}$ and $\ket{\mu_2}$ have different geometric-like measure of imaginarity. We mix them with a rotation angle $\alpha \in [0,\pi/2]$,
\begin{eqnarray}
	\sqrt{\lambda_1'}\ket{\mu'_1} &= \cos{\alpha}\sqrt{\lambda_1}\ket{\mu_1} + \sin{\alpha}\sqrt{\lambda_{2}}\ket{\mu_2},\nonumber\\
	\sqrt{\lambda_2'}\ket{\mu'_2} &= -\sin{\alpha}\sqrt{\lambda_{1}}\ket{\mu_1} + \cos{\alpha}\sqrt{\lambda_{2}}\ket{\mu_2}.\nonumber
\end{eqnarray}
By continuously varying $\alpha$ one obtains that $\mathcal{M}_{gl}(\ket{\mu'_1}) =\mathcal{M}_{gl}({\rho})$. This process can be recursively applied to the remaining pure states until the desired decomposition is obtained, while $\sum_i \lambda_i' \braket{\mu_i'}{\mu'^{*}_i}=1-\mathcal{M}_{gl}(\rho)$ is kept.
$\hfill\blacksquare$

Analogous to two-qubit entanglement theory \cite{Vidal22,Wei36,Wootters377}, Lemma 1 proves the existence of a pure state decomposition with identical geometric-like measure of imaginarity. This allows us to derive the following Lemma.

{\bf Lemma 2}.\label{GeomContinuity}
For an arbitrary state $\rho$, consider the set $S_{\rho,f}$ containing states $\rho'$ such that $F(\rho,\rho')\geq f$ for all $\rho' \in S_{\rho,f}$. The minimal geometric-like measure of imaginarity within $S_{\rho,f}$ is given by
\begin{eqnarray*}
	\min_{\rho'\in S_{\rho,f}}\mathcal{M}_{gl}(\rho')=1-\cos\left(\max\left\{\Delta_1, 0\right\} \right),
\end{eqnarray*}
where $\Delta_1=\sin^{-1}\left(\sqrt{2\mathcal{M}_{gl}(\rho)+\mathcal{M}^2_{gl}(\rho)}\right)-\cos^{-1}(\sqrt{f})$.
For any pure state $\ket{\psi}$, the maximal geometric-like measure of imaginarity in $S_{\ket{\psi},f}$ is given by
\begin{eqnarray*}
	\max_{\rho'\in S_{\ket{\psi},f}}\mathcal{M}_{gl}(\rho')=1-\cos\left(\min\left\{\Delta_2, \frac{\pi}{4}\right\} \right),
\end{eqnarray*}
where $\Delta_2=\sin^{-1}\left(\sqrt2\mathcal{M}_{gl}(\ket{\psi})+\mathcal{M}^2_{gl}(\ket{\psi})\right) +\cos^{-1}(\sqrt{f})$.

{\sf Proof.}
We employ the Bures angle as a distance measure between states of $\rho$ and $\sigma$, $D(\rho, \sigma) = \cos^{-1}\left(\sqrt{F(\rho, \sigma)}\right)$.
Note that the geometric-like measure of imaginarity $\mathcal{M}_{gl}(\rho)$ falls within the range of $0$ to $\frac{1}{2}$. Consider two states, $\rho$ and $\rho'$, where $\rho'$ belongs to the set $S_{\rho,f}$. Assume that $\rho_r$ and $\rho_r'$ are the closest real states closest to $\rho$ and $\rho'$, respectively, as measured by the Bures angle. From the fact that $\rho_r$ is the closest real state to $\rho$, employing the triangle inequality of distances we derive the following chain inequalities,
\begin{eqnarray}
	\cos^{-1}\left(1-\mathcal{M}_{gl}(\rho)\right)
	&=&D(\rho,\rho_r) \leq D(\rho,\rho'_r) \nonumber\\
	&\leq& D(\rho,\rho')+D(\rho', \rho'_r) \nonumber\\
	&\leq& \cos^{-1}(\sqrt{f}) + \cos^{-1}\left(1-\mathcal{M}_{gl}(\rho')\right).\nonumber
\end{eqnarray}
Hence,
\begin{eqnarray} \label{geometric_lower_bound}
	\mathcal{M}_{gl}(\rho') \geq 1-\cos\left(\max\left\{\Delta_1, 0\right\} \right).
\end{eqnarray}

In a similar way, from the fact that $\rho'_r$ is the real state closest to $\rho'$ we obtain
\begin{align}
	\cos^{-1}\left(1-\mathcal{M}_{gl}(\rho')\right)
	&=D(\rho',\rho'_r)\leq D(\rho',\rho_r)\nonumber\\
	&\leq D(\rho,\rho')+D(\rho, \rho_r)\nonumber\\
	&\leq \cos^{-1}(\sqrt{f}) + \cos^{-1}\left(1-\mathcal{M}_{gl}(\rho)\right)\nonumber.
\end{align}
Therefore, we arrive at
\begin{eqnarray} \label{geometric_upper_bound}
	\mathcal{M}_{gl}(\rho') \leq 1-\cos\left(\min\left\{\Delta_2, \frac{\pi}{4}\right\} \right).
\end{eqnarray}

We now demonstrate that the lower bound in (\ref{geometric_lower_bound}) can indeed be achieved. Based on Lemma 1, for any given state $\rho$, there exists a pure state decomposition $\rho = \sum_i p_i \ket{\psi_i}\!\bra{\psi_i}$ such that $\mathcal{M}_{gl}(\ket{\psi_i}) =\mathcal{M}_{gl}(\rho)$ for all pure states $\ket{\psi_i}$. This decomposition implies that each pure state $\ket{\psi_i}$ can be expressed as
$\ket{\psi_i}= \cos{\alpha}\ket{a_i} + i \sin{\alpha}\ket{a^{\perp}_i}$,
where $\braket{a_i}{a^{\perp}_i} = 0$, and $\ket{a_i}$ and $\ket{a^{\perp}_i}$ are real states. Here, we set $\alpha=\sin^{-1}\left(\sqrt{2\mathcal{M}_{gl}(\rho)+\mathcal{M}^2_{gl}(\rho)}\right)$, where $\alpha \in\{0,\frac{\pi}{4}\}$. 

Consider the state $\rho_{\min} = \sum_i q_i \ket{\phi_i}\bra{\phi_i}$, where
$\ket{\phi_i}= \cos{\tilde{\beta}}\ket{a_i}  + \sin{\tilde{\beta}}\ket{a^{\perp}_i}$, $\tilde \beta \in [0,\pi/4]$ and $q_i = \frac{p_i |\braket{\psi_i}{\phi_i}|^2}{\sum_k p_k |\braket{\psi_k}{\phi_k}|^2}$.
Utilizing the convexity property of the geometric-like measure of imaginarity, we obtain
\begin{eqnarray}
	\mathcal{M}_{gl}(\rho_{\min})\leq 1-\cos\tilde{\beta}. \label{convexity}
\end{eqnarray}
From the properties of the root-fidelity \cite{Wilde2017}, we have
\begin{eqnarray}
	\sqrt{F(\rho, \rho_{\min})} &\geq& \sum_i \sqrt{p_i q_i} |\braket{\psi_i}{\phi_i}| \nonumber \\
	&=& \sqrt{\sum_i p_i |\braket{\psi_i}{\phi_i}|^2}\nonumber\\
	&=& |\cos (\alpha -\tilde{\beta})|.\nonumber
\end{eqnarray}
Therefore, $F(\rho, \rho_{\min}) \geq \cos^{2}(\alpha -\tilde{\beta})$.
For the sake of convenience, we set
$\tilde{\beta}=\max\{\alpha-k,0\}$ and $k=\cos^{-1}\sqrt{f}$.
Then
$$
F(\rho, \rho_{\min}) \geq \cos^{2}\left(\min\{k,\alpha\}\right) \geq \cos^{2}k\geq f. 
$$
Here, we have taken into account that the cosine squared function is monotonically decreasing in the interval $[0, \pi/4]$, while the value of $\min\{k,\alpha\}$ is within this interval. This shows that the state $\rho_{\min}$ is within $S_{\rho,f}$. Furthermore, using (\ref{convexity}) we obtain
\begin{align}
	\mathcal{M}_{gl}(\rho_{\min})
	\leq&1-\cos\Big(\max\{\alpha-k,0\}\Big)\nonumber\\
	\leq&1-\cos\Big(\max\Big\{\sin^{-1}(\sqrt{2\mathcal{M}_{gl}(\rho)+\mathcal{M}^2_{gl}(\rho)})\nonumber\\
	&-\cos^{-1}(\sqrt{f}),0\Big\}\Big). \label{eq:btildeUpperBound}
\end{align}
Comparing (\ref{geometric_lower_bound}) with (\ref{eq:btildeUpperBound}), we get
\begin{align}
	\mathcal{M}_{gl}(\rho_{\min}) =&1-\cos\Big(\max\Big\{\sin^{-1}(\sqrt{2\mathcal{M}_{gl}(\rho)+\mathcal{M}^2_{gl}(\rho)})\nonumber\\
	&-\cos^{-1}(\sqrt{f}),0\Big\}\Big).\nonumber
\end{align}
Hence, the minimum geometric-like measure of imaginarity within the set $S_{\rho,f}$ is given by
\begin{align}\label{min_geo}
	\min_{\rho'\in S_{\rho,f}}\mathcal{M}_{gl}(\rho') 
	=&1-\cos\Big(\max\Big\{\sin^{-1}(\sqrt{2\mathcal{M}_{gl}(\rho)+\mathcal{M}^2_{gl}(\rho)})\nonumber\\
	&-\cos^{-1}(\sqrt{f}),0\Big\}\Big).
\end{align}
$\bar{f}$
Now, we consider the case where $\rho$ is a pure state $\ket{\psi}=\cos{\alpha}\ket{a} + i \sin{\alpha}\ket{a^{\perp}}$. We demonstrate that the upper bound specified in (\ref{geometric_upper_bound}) is also attainable. To this end, let us consider
$\ket{\psi_{\max}}=\cos(\min \{ \alpha + k, \pi/4\})\ket{a}+ i \sin(\min \{ \alpha + k, \pi/4\})\ket{a^{\perp}}$.
The geometric-like measure of imaginarity for $\ket{\psi_{\max}}$ is given by
$\mathcal{M}_{gl}(\ket{\psi_{\max}})=1-\cos(\min \{ \alpha + k, \pi/4\})$.
The fidelity between $\psi$ and $\psi_{\max}$ satisfies the following relation,
\begin{eqnarray}
	F(\psi, \psi_{\max}) = \cos^{2}\left(\min \left\{k,\frac{\pi}{4}-\alpha\right\} \right) \geq \cos^{2}k\geq f,\nonumber
\end{eqnarray}
as $\cos^2$ is a monotonically decreasing function in $[0, \pi/4]$ and $\min \{k,\pi/4-\alpha\}$ is within $[0, \pi/4]$. Consequently, $\ket{\psi_{\max}}$ belongs to the set $S_{\ket{\psi},f}$ and attains the maximum possible geometric-like measure of imaginarity within this set. Specifically, we have
\begin{align}
	\max_{\rho'\in S_{\ket{\psi},f}}\mathcal{M}_{gl}(\rho') =&1-\cos\Big(\min\{\alpha+k,\pi/4\}\Big)\nonumber\\
	=&1-\cos\Big(\min\Big\{\sin^{-1}(\sqrt{2\mathcal{M}_{gl}(\rho)+\mathcal{M}^2_{gl}(\rho)})\nonumber\\
	&+\cos^{-1}\sqrt{f},\pi/4\Big\}\Big).\nonumber
\end{align}
This completes the proof of the attainability of both the lower and upper bounds on the geometric-like measure of imaginarity within the fidelity-constrained set.
$\hfill\blacksquare$

Lemma 2 provides the minimum state $\rho'$ such that the geometric-like measure of imaginarity between $\rho'$ and $\rho$ is at least the target fidelity $f$.

{\bf Theorem 4}. \label{PureConversion2}
The maximal probability of stochastic-approximate state transformations from a pure state $\ket{\psi}$ to an arbitrary state $\rho$ via real operations with fidelity $f$ is given by
\begin{eqnarray}
	P_{f}(\ket{\psi}\rightarrow\rho) = \begin{cases}
		1,~~m_1\geq0\\[2mm]
		\frac{\mathcal{M}_{gl}(\ket{\psi})}{1-\cos(\Delta_1)},~~\mathrm{otherwise},
	\end{cases}\nonumber
\end{eqnarray}
where\\ $m_1=\sin^{-1}\sqrt{2\mathcal{M}_{gl}(\ket{\psi})+\mathcal{M}^2_{gl}(\ket{\psi})}-\sin^{-1}\sqrt{2\mathcal{M}_{gl}(\rho)+\mathcal{M}^2_{gl}(\rho)} +\cos^{-1}\sqrt{f}$,\\
$\Delta_1=\sin^{-1}\Big(\sqrt{2\mathcal{M}_{gl}(\rho)+\mathcal{M}^2_{gl}(\rho)}\Big)-\cos^{-1}\sqrt{f}$.

{\sf Proof.} 
To obtain state $\rho$ with a fidelity of at least $f$, the optimal strategy is to transfer to a state $\rho'$ in $S_{\rho,f}$ with a minimal geometric-like measure of imaginarity,
\begin{eqnarray}
	P_{f}(\ket{\psi}\rightarrow\rho)=\min\left\{\frac{\mathcal{M}_{gl}(\ket{\psi})}{\min_{\rho'\in S_{\rho,f}}\mathcal{M}_{gl}(\rho')},1\right\}.\nonumber
\end{eqnarray}
Using Eq. (\ref{min_geo}) we have
\begin{eqnarray}
	\min_{\rho'\in S_{\rho,f}}\mathcal{M}_{gl}(\rho')
	&=&1-\cos\Big(\max\Big\{\sin^{-1}(\sqrt{2\mathcal{M}_{gl}(\rho)+\mathcal{M}^2_{gl}(\rho)})\nonumber\\
	&&-\cos^{-1}(\sqrt{f}),0\Big\}\Big).\nonumber
\end{eqnarray}

Denote
\begin{eqnarray}
	m_1=&&\sin^{-1}\sqrt{2\mathcal{M}_{gl}(\ket{\psi})+\mathcal{M}^2_{gl}(\ket{\psi})}\nonumber\\
	&&-\sin^{-1}\sqrt{2\mathcal{M}_{gl}(\rho)+\mathcal{M}^2_{gl}(\rho)} +\cos^{-1}\sqrt{f}.\nonumber
\end{eqnarray}
When $m_1\geq0$, we obtain
\begin{eqnarray}
	&&\sin^{-1}\sqrt{2\mathcal{M}_{gl}(\ket{\psi})+\mathcal{M}^2_{gl}(\ket{\psi})}\nonumber\\
	&&\geq\sin^{-1}\sqrt{2\mathcal{M}_{gl}(\rho)+\mathcal{M}^2_{gl}(\rho)} -\cos^{-1}\sqrt{f}.\nonumber
\end{eqnarray}
Given that $\sin^{-1}\sqrt{2\mathcal{M}_{gl}(\rho)+\mathcal{M}^2_{gl}(\rho)} -\cos^{-1}\sqrt{f}\in[-\frac{\pi}{2},\frac{\pi}{4}]$ and $\sin^{-1}\sqrt{\mathcal{M}_{gl}(\ket{\psi})}\in[0,\frac{\pi}{4}]$,
we have
\begin{eqnarray}
	&&\sin^{-1}\sqrt{2\mathcal{M}_{gl}(\ket{\psi})+\mathcal{M}^2_{gl}(\ket{\psi})}\nonumber\\
	&&\geq\max\left\{\sin^{-1}\sqrt{2\mathcal{M}_{gl}(\rho)+\mathcal{M}^2_{gl}(\rho)} -\cos^{-1}\sqrt{f}, 0 \right\}.\nonumber
\end{eqnarray}
Using these relations, we obtain
\begin{align}
	\min_{\rho'\in S_{\rho,f}}\mathcal{M}_{gl}(\rho')
	=&1-\cos\Big(\max\Big\{\sin^{-1}\sqrt{2\mathcal{M}_{gl}(\rho)+\mathcal{M}^2_{gl}(\rho)}\nonumber\\
	& -\cos^{-1}\sqrt{f}, 0 \Big\} \Big)\nonumber\\
	\leq& 1-\cos\left(\sin^{-1}\sqrt{2\mathcal{M}_{gl}(\ket{\psi})+\mathcal{M}^2_{gl}(\ket{\psi})}\right)\nonumber\\
	=&\mathcal{M}_{gl}(\ket{\psi}).\nonumber
\end{align}
When $\mathcal{M}_{gl}(\ket{\psi}) > 0$, it follows that
\begin{eqnarray}
	\frac{\mathcal{M}_{gl}(\ket{\psi})}{\min_{\rho'\in S_{\rho,f}}\mathcal{M}_{gl}(\rho')} \geq 1,\nonumber
\end{eqnarray}
which implies that $P_{f}(\ket{\psi}\rightarrow\rho)=1$ for $m_1\geq 0 $.

For the case of $m_1<0$, we have
\begin{eqnarray}
&&\sin^{-1}\sqrt{2\mathcal{M}_{gl}(\rho)+\mathcal{M}^2_{gl}(\rho)}-\cos^{-1}\sqrt{f}\nonumber\\
&&>\sin^{-1}\sqrt{2\mathcal{M}_{gl}(\ket{\psi})+\mathcal{M}^2_{gl}(\ket{\psi})}\nonumber\\
&&>0.\nonumber
\end{eqnarray}
Using Eq. (\ref{optprob}), we obtain
\begin{eqnarray}
&&P_{f}(\ket{\psi}\rightarrow\rho)\nonumber\\
=&&\frac{\mathcal{M}_{gl}(\ket{\psi})}{1-\cos(\sin^{-1}\sqrt{2\mathcal{M}_{gl}(\rho)+\mathcal{M}^2_{gl}(\rho)}-\cos^{-1}\sqrt{f})}.\nonumber
\end{eqnarray}
This completes the proof.
$\hfill\blacksquare$

Theorem 4 shows that the maximal probability of stochastic-approximate state transformations from a pure state $\ket{\psi}$ to an arbitrary state $\rho$ via real operations with fidelity $f$ is also characterized by the geometric-like measure of imaginarity.

\section{Conclusions}
The resource-theoretic approach has recently gained prominence in quantum mechanics and quantum information theory, offering a rigorous mathematical framework to elucidate the significance of complex numbers in this field and enabling a more systematic and quantitative examination of complex numbers in quantum systems. Additionally, there has been a surge of interest in quantifying imaginarity quantities, particularly within the context of entanglement and coherence theory. The geometric-like measure of imaginarity introduced in this study particularly shows its practical applications in characterizing quantum state transformations.

Compared to the geometric imaginarity measure, our geometric-like measure of imaginarity exhibits advantages including greater robustness against noise and decoherence, higher stability and less information loss during transmission. Furthermore, in state conversion problems, this measure simplifies the computation of transformation probabilities while ensuring maximum fidelity of the transformed state. Our study highlights the profound importance of complex numbers in quantum physics and the further investigations on the applications of the imaginarity in quantum information processing.

\section*{ACKNOWLEDGMENTS}
This work is supported by the National Natural Science Foundation of China (NSFC) under Grants 12075159, 12171044 and 12175147; the specific research fund of the Innovation Platform for Academicians of Hainan Province.

\end{document}